\documentclass{llncs}

\usepackage{amsfonts,amsmath,amssymb}
\usepackage[noend]{algorithmic}
\usepackage{algorithm}
\usepackage{graphicx}
\usepackage{listings}
\usepackage{multirow}

\lstdefinelanguage{pseudocode}{
morekeywords={if, then, else, for, return, in, while, case, do, forever, foreach, False, True},
morecomment=[l]\#,%
morecomment=[l]//,%
moredelim=[il][\it]{|},
moredelim=[il][]{[},escapeinside={(*@}{@*)}
}
\begin{document}

\title{Chaining fragments in sequences: to sweep or not}

\author{Julien Allali\inst{1,2} \and
  Cedric Chauve\inst{2,3} \and
  Laetitia Bourgeade\inst{1}}

\institute{
  LaBRI, Universit\'e Bordeaux, France \and
  IBP, Universit\'e Bordeaux, France \and
  Department of Mathematics, Simon Fraser University, Canada}

\maketitle{}

\begin{abstract} 
  Computing an optimal chain of fragments is a classical problem in string
  algorithms, with important applications in computational
  biology. There exist two efficient dynamic programming algorithms
  solving this problem, based on different principles. In the present
  note, we show how it is possible to combine the principles of two of
  these algorithms in order to design a hybrid dynamic programming
  algorithm that combines the advantages of both algorithms.
\end{abstract}

\section{Introduction}\label{sec:intro}

The need for very efficient pairwise sequence alignments algorithm has
motivated the development of methods aimed at breaking the natural
quadratic time complexity barrier of dynamic programming alignment
algorithms \cite{Gusfield-algorithms}. One of the successful
alternative approaches is based on the technique of chaining
fragments. Its principle is to first detect and score highly conserved
factors, the {\it fragments} (also called {\it anchors} or {\it
  seeds}), then to compute a maximal score subset of fragments that
are colinear and non-overlapping in both considered sequences, called
an {\it optimal chain}. This optimal chain is then used as the
backbone of an alignment, that is completed in a final stage by
aligning the gaps located between consecutive selected fragments. This
approach is used in several computational biology applications, such
as whole genome comparison~\cite{UricaruMR11,AbouelhodaO05,MGA},
cDNA/EST mapping~\cite{Ohlebusch-chaining}, or identifying regions
with conserved synteny.

In the present work we are interested in the problem of computing an
optimal chain of fragments\footnote{We focus here on the problem of
  computing the score of an optimal chain, but our algorithm can be
  complemented by a standard backtracking procedure to compute an
  actual optimal chain.}, from a given set of $k$ fragments, for two
sequences $t$ and $u$ of respective lengths $n$ and $m$. Due to its
applications, especially in computational biology, this problem has
received a lot attention from the algorithmic
community~\cite{Eppstein-sparse,Felsner-trapezoid,Joseph-determining,Morgenstern-simple,Myers-chaining,Myers-O,AbouelhodaO05}.
The fragment chaining problem can be solved in $O(k+n\times m)$ time
by using a simple dynamic programming (DP) algorithm
(see~\cite{Morgenstern-simple} for example).  However, in practical
applications, the number $k$ of fragments can be subquadratic, which
motivated the design of algorithms whose complexity depends only of
$k$ and can run in $O(k \log k)$ worst-case time
(see~\cite{Joseph-determining,Felsner-trapezoid,Myers-chaining,Ohlebusch-chaining}). The
later algorithms, known as Line Sweep (LS) algorithms, rely on
geometric properties of the problem, where fragments can be seen as
rectangles in the quarter plane, and geometric data structures that
allow to retrieve and update efficiently (\emph{i.e.} in logarithmic
time) optimal subchains(see~\cite{Ohlebusch-chaining} for example).

This raises the natural question of deciding which algorithm to use to
when comparing two sequences $t$ and $u$. In particular, it can happen
that the {\it density} of fragments differs depending on the location
of the fragments in the considered sequences, due for example for the
presence in repeats. In such cases, it might then be more efficient to
rely on the DP algorithm in regions with high fragment density,
while in regions of lower fragment density, the LS algorithm would be
more efficient. This motivates the theoretical question we consider,
that asks to design an efficient algorithm that relies on the
classical DP principle when the density of fragments is high and
switches to the LS principle when processing parts of the sequences
with a low density of fragments. We show that this can be achieved,
and we describe such a \emph{hybrid} DP/LS algorithm for computing the
score of an optimal chain of fragments between two sequences. We prove
that our algorithm achieve a theoretical complexity that is as good as
both the DP and LS algorithm, \emph{i.e.} that for any instance, our
algorithm performs as at least as well, in terms of theoretical
worst-case asymptotic time complexity, as both the DP and the LS
algorithm.

In Section \ref{sec:prelim}, we introduce formally the fragment
chaining problem and the DP and LS algorithms. In Section
\ref{sec:results}, we describe our hybrid algorithm and analyze its
complexity.

\section{Preliminaries}\label{sec:prelim}

\paragraph{Preliminary definitions and problem statement.}
Let $t$ and $u$ be two sequences, of respective lengths $n$ and
$m$. We assume that positions index in sequences start at $0$, so
$t[0]$ is the first symbol in $t$ and $t[n-1]$ its last symbol. As
usual, by $t[i,j]$ we denote the substring of $t$ composed of symbols
in positions $i,i+1,\dots,j$.

\begin{figure}[!htb]
  \begin{center}
    \includegraphics[width=0.6\linewidth]{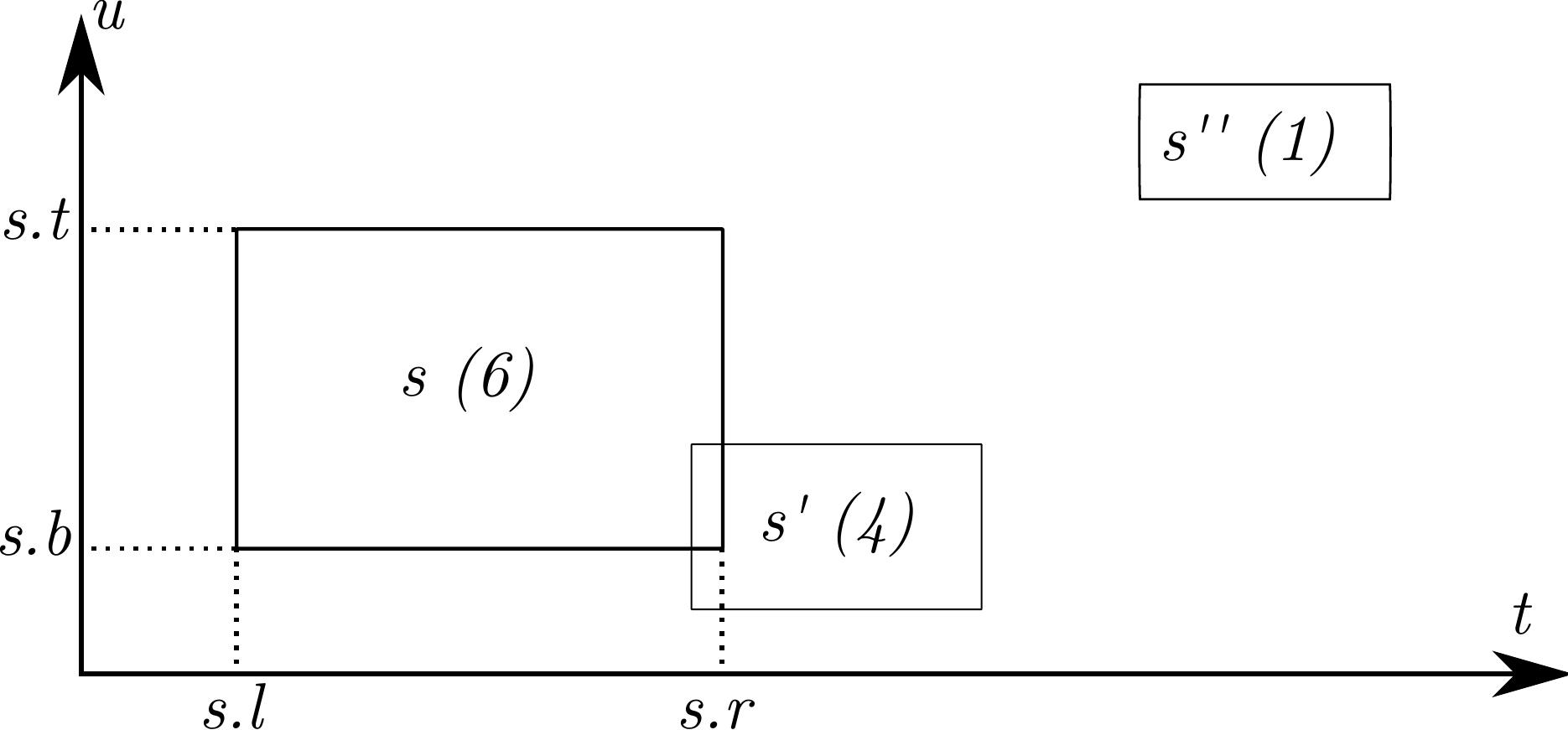}
    \caption{Example of the fragment chaining problem with three
      fragments represented by squares. Possible chains are
      $[(s),(s'),(s''),(s,s''),(s',s'')]$. The best chain is
      $(s,s'')$, with a score of $7$.}
    \label{fig:example}
  \end{center}
\end{figure}

A {\it fragment} is a factor that is common, possibly up to small
variations, to $t$ and $u$. Formally, a fragment $s$ is defined by 5
elements $(s.\ell,s.r,s.t,s.b,s.s)$: the first four fields indicate
that the corresponding substrings are $t[s.\ell,s.r]$ and
$u[s.b,s.t]$, while the field $s.s$ is a {\it score} associated to the
fragment.  We call $borders$ of $s$ the coordinates $(s.{\ell},s.b)$
and $(s.r,s.t)$.  As usual in chaining problems, we see fragments as
rectangles in the quarter plane, where the $x$-axis corresponds to $t$
and the $y$-axis to $u$. For a fragment $s$, $s.{\ell}$,$s.r$,$s.b$
and $s.t$ denote the $left$ and $right$ position of $s$ over $t$ and
the $bottom$ and $top$ position of $s$ over $u$ ($s.{\ell} \le s.r$
and $s.b \le s.t$). See figure~\ref{fig:example} for an example.

Let ${\cal S}$ denote a set of $k$ fragments for $t$ and $u$.  A {\it
  chain} is a set of fragments $\{s_1,\dots,s_\ell\}$ such that $s_i.r
< s_{i+1}.{\ell}$ and $s_i.t < s_{i+1}.b$ for $i=1,\dots,\ell-1$; the
score of a chain is the sum $\sum_{i=1}^\ell s_i.s$ of the fragments
it contains.  A chain is optimal if there is no chain with a higher
score. The problem we consider in the present work is to compute the
score of an optimal chain.

\paragraph{The dynamic programming (DP) algorithm.} 
We first present a simple dynamic programming (DP) algorithm that
computes an $n\times m$ dynamic programming table $M$ such that
$M[i][j]$ is the score of an optimal chain for the prefixes $t[0,i]$
and $u[0,j]$ (See pseudo-code~\ref{algo:dynprog}).  We present here a
version that does not instantiate the full $n\times m$ DP table, but
records only the last filled column, following the classical technique
introduced in~\cite{Hirschberg-linear} and used in the space-efficient
fragment chaining DP algorithm described in~\cite{Morgenstern-simple}.

\begin{algorithm}\small
\caption{The Dynamic Programming algorithm}
\label{algo:dynprog}
\begin{lstlisting}[]
$L$: an array of $n\times m$ linked lists
$S$: an array of $k$ integers
$M$: an array of $m$ integers
foreach $s$ in ${\cal S}$ do
   front insert $(s,end)$ into  $L[s.r][s.t]$
   front insert $(s,begin)$ into $L[s.{\ell}][s.b]$
for $i$ from $0$ to $n$ (*@\label{algo:dp:column_loop}@*)
   $left = 0$
   $leftDown = 0$
   for $j$ from $0$ to $m$ (*@\label{algo:dp:row_loop}@*)
      $maxC = 0$
      foreach $(s,type)$ in $L[i][j]$
         if $type$ is $begin$
            $S[s] = s.s + leftDown$
         if $type$ is $end$ and $S[s]>maxC$
            $maxC = S[s]$
      $leftDown = left$
      $left = M[j]$
      $M[j] = max(M[j],M[j-1],maxC)$
return $M[m-1]$
\end{lstlisting}
\end{algorithm}

The difference with Morgenstern's space efficient DP
algorithm~\cite{Morgenstern-simple} is that we still require a
quadratic space for the data structure $L$. In terms of computing the
score of an optimal chain, the key point is that $S[s]$, if defined,
contains the optimal score of a chain that contains $s$ as last
fragment.  The worst-case time complexity of this algorithm is
obviously $O(k + n\times m)$.

\paragraph{The Line Sweep (LS) algorithm.} 
We now describe a Line Sweep algorithm for the fragment chaining
problem (See pseudo-code~\ref{algo:linesweep}). The main idea is to
process fragments according to their order in the sequence $t$, while
maintaining a data structure that records, for each position $i$ in
$u$, the best partial chain found so far using only fragments below
position $i$.

\begin{algorithm}\small
\caption{The Line Sweep algorithm}
\label{algo:linesweep}
\begin{lstlisting}[]
$P$: an array of $2k$ triples $(position,type,fragment)$
$A$: a set of pairs $(position,score)$ 
$S$: an array of $k$ integers
foreach $s$ in ${\cal S}$ do
   insert $(s.{\ell},begin,s)$ into $P$
   insert $(s.r,end,s)$ into $P$
sort $P$ according to the field position, with $begin$ positions appearing before $end$ positions having the same value
foreach $(pos,type,s)$ in $P$ (*@\label{algo:ls:process_end}@*)
    if $type$ is $begin$
        retrieve from $A$ the pair $(p,v)$ such that $p$ is the highest position strictly less than $s.b$
        $S[s] = s.s + v$ (*@\label{algo:ls:use_A}@*)
    if $type$ is $end$
        $(p,v)$ = retrieve from $A$ the highest position less or equal to $s.t$ (*@\label{algo:ls:A_1}@*)
        if $S[s]>v$
           retrieve from $A$ the pair $(p',v')$ such that $v'$ is the highest score less than or equal to $S[s]$
           remove  from $A$ all entries $(p'',v'')$ such that $p<p''\leq p'$ (*@\label{algo:ls:update_A}@*)
           insert $(s.t,S[s])$ into $A$  (*@\label{algo:ls:A_2}@*)
$(p,v)$ = last entry of $A$
return $v$
\end{lstlisting}
\end{algorithm}

In this algorithm $P$ stores all fragments borders, $S[s]$, as in the
DP algorithm, is the score of an optimal chain among all the chains
that end with fragment $s$. A fragment $s$ is said to have been {\it
  processed} after the entry $(s.r,end,s.s)$ has been processed
through the loop in line \ref{algo:ls:process_end}. A {\it partial
  chain} is a chain composed only of processed fragments.

The data structure $A$ satisfies the following invariant, that is key
to ensure the correctness of the algorithm: if $(pos,type,s)$ is the
last entry of $P$ that has been processed, then $A$ contains an entry
$(p,v)$ if and only if the best chaining score, among partial chains
that belong to the rectangle defined by points $(0,0)$ and $(pos,p)$,
is $v$ and corresponds to a chain ending with a fragment $s'$ such
that $s'.t=p$. 

Line \ref{algo:ls:update_A} ensures this invariant is maintained. This
invariant allows to retrieve from $A$ the score of an optimal partial
chain that can be extended by the current fragment $s$, \emph{i.e.} that ends
up in $u$ in a position strictly smaller than $s.b$ (line
\ref{algo:ls:use_A}). This property follows from the fact that the
order in which fragments are processed ensures that all previously
processed fragments do not overlap with the current fragment in $t$.

In order to implement this algorithm efficiently, it is fundamental to
ensure that in line \ref{algo:ls:update_A}, the time required to
remove $c$ entries (the set of all entries of $A$ with first field
strictly greater than $p$ and lower than or equal to $p'$) is
$O(c\log(k))$. If $A$ is implemented in a data structure that
satisfies this property and support searches, insertions and deletions
in logarithmic time, then the time complexity of the algorithm is $O(k
\log(k))$; see \cite{Ohlebusch-chaining} for a discussion on such data
structures.

\section{An hybrid algorithm}\label{sec:results}

We now describe an algorithm that combines both approaches described
in the previous section.  

\paragraph{Overview.} 
We first introduce the notion of {\it compact instance}.  An instance
of the chaining problem is said to be compact, if each position of $t$
and each position of $u$ contains at least one border. If an instance
is not compact, then there exists a unique compact instance obtained
by removing from $t$ and from $u$ all positions that do not contain a
fragment border, leading to sequences $t'$ and $u'$, and updating the
fragments borders according to the sequences $t'$ and $u'$, leading to
a set ${\cal S}'$ of fragments. From now, we denote by $(t',u',{\cal
  S}')$ the compact instance corresponding to $(t,u,{\cal S})$, and
$m'$ and $n'$ the lengths of $t'$ and $u'$.

Next, we define, for a position $p$ of $t$ its {\it border density}
${\cal K}_p$ as the number of fragment borders (\emph{i.e.} number of fragments
extremities) located in $t[p]$. If $P^1$ is the set of positions in
$t'$ with border density strictly greater than $\frac{m'}{\log m'
  -1}$, and $P^2$ the remaining $n'-|P^1|$ positions of $t'$, then the
hybrid DP/LS algorithm we describe below has time complexity

$$O\left( k + \min(k\log(k),m) + \min(k\log(k),n) + \sum_{p \in P^1} (
m' + \mathcal{K}_p ) + \log(m') \sum_{p \in P^2}
\mathcal{K}_p\right).$$

Intuitively, our hybrid algorithm works on a compact instance, and
fills in the DP table for this compact instance, deciding for each
column of this table (\emph{i.e.} position of $t'$) to fill it in using the
DP equations or the Line Sweep principle, based on its border density.

\paragraph{{Compacting an instance}\label{sec:compact}.}
We first describe how to compute the compact instance $(t',u',{\cal
  S}')$.

\begin{lemma}\label{lem:compact}
  The compact instance $(t',u',{\cal S}')$ can be computed in time
  \\ $O\left( k + \min(k\log(k),m) + \min(k\log(k),n)\right)$ and
  space $O(k+n+m)$.
\end{lemma}

The proof of this lemma is quite straightforward, and we omit the
details here for space reason. Assume we are dealing with $t$ (the
same method applies to $u$).
\begin{itemize}
\item If $k\log (k)\leq m$, then we (1) sort the fragments extremities
  in $t$ in increasing order of their starting position, (2) cluster
  together fragment extremities with the same value, and (3) relabel
  the coordinates of each fragment extremity using the number of
  clusters preceding it in the order, plus one.
\item If $k\log k > m$, then we (1) detect positions of $t$ with no
  fragment extremities, in $O(k+m)$ time, (2) mark them and relabel
  the positions with non-zero density in $O(m)$ time, and finally (3)
  relabel the fragment extremities according to the new labels of
  their positions, in $O(k)$ time.
\end{itemize}

From now, we assume that the compact instance has been computed and
that it is the considered instance.

\paragraph{DP update vs LS update.}

In this section, we introduce our main idea. The principle is to
consider fragments in the same order than in the LS algorithm --
\emph{i.e.} through a loop through indices $0$ to $n'-1$, a feature which is
common to both the DP and LS algorithms --, but to process the
fragments whose border in $t'$ is in position $i$ using either the DP
approach if the density of fragments at $t'[i]$ is high, or the LS
approach otherwise.  Hence, the key requirement will be that,
\begin{itemize}
\item when using the DP approach, the previous column of the DP table
  is available,
\item when using the LS approach, a data structure with similar
  properties than data structure $A$ used in the LS algorithm is
  available.
\end{itemize}

\paragraph{A hybrid data structure.} 
We introduce now a data structure $B$ that ensures that the above
requirements are satisfied.  The data structure $B$ is essentially an
array of $m'$ entries augmented with a balanced binary search
tree. Formally:
\begin{itemize}
\item We consider an array $\mathcal{B}$ of $m'$ entries, such that
  $\mathcal{B}[i]$ contains chaining scores, and satisfies the
  following invariant: if $s$ is the last processed fragment, for
  every $i=1,\dots,s.r$, $\mathcal{B}[i]\geq \mathcal{B}[i-1]$.
\item
  We augment this array with a balanced binary search tree
  $\mathcal{C}$ whose leaves are the entries of $\mathcal{B}$ and whose
  internal nodes are labeled in order to satisfy the following
  invariant: a node $x$ is labeled by the maximum of the labels of its
  right child and left child.
\end{itemize}

The data structure $B$ will be used in a similar way than the data
structure $A$ of the LS algorithm, \emph{i.e.} to answer the following
queries: given $0\leq p \leq m'$, find the optimal score of a partial
chain whose last fragment $s$ satisfies $s.t\leq p$. This principle is
very similar to solutions recently proposed for handling dynamic
minimum range query requests~\cite{Arge-dynamic}.

\begin{figure}
  \begin{center}
    \includegraphics[width=0.8\linewidth]{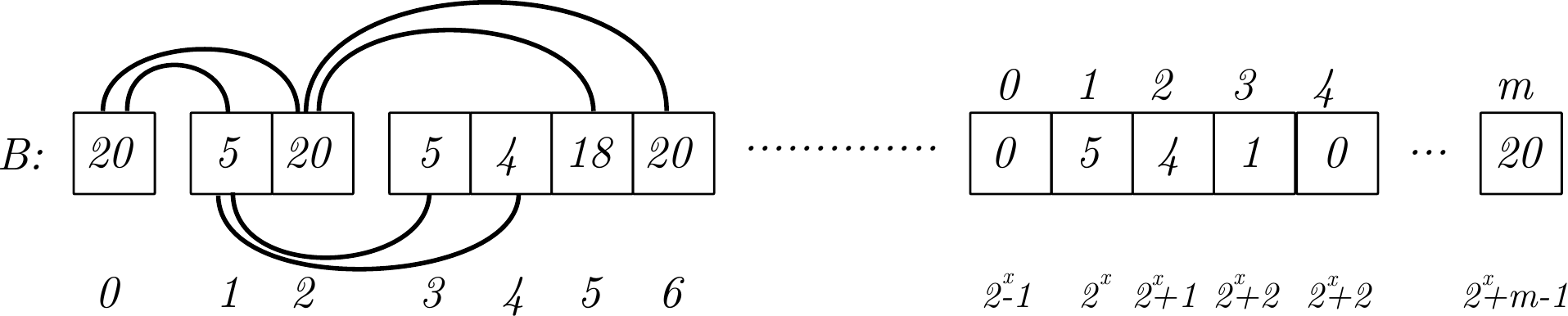}
    \caption{Example of the implementation of the data structure $B$
      with an array.}
    \label{fig:bintree}
  \end{center}
\end{figure}

We describe now how we implement this data structure using an array.
Let $b$ be the smallest integer such that $m'\leq 2^b$. We encode $B$
into an array of size $2^{b+1}$, whose prefix of length $m'-1$
contains the labels of the internal nodes of the binary tree
$\mathcal{C}$ (so each cell contains a label and the indexes to two
other cells, corresponding respectively to the left child and right
child), ordered in breadth-first order, while the entries of
$\mathcal{B}$ are stored in the suffix of length $m'$ of the array
(see figure~\ref{fig:bintree}). From now, we identify nodes of the
binary tree and cells of the array, that we denote by $B$.

Using this implementation, for a given node of the binary search tree,
say encoded by the cell in position $x$ in $B$ (called node $x$ from
now), we can quickly obtain the position, in the array, of its left
child, of its right child, but also of its parent (if $B[x]$ is not
the root) and of its rightmost descendant, defined as the unique node
reached by a maximal path of edges to right children, starting at $x$
edges to a left (resp. right) child.  Indeed, it is straightforward to
verify that, the constraint of ordering the nodes of the binary tree
in the array according to a breadth-first order implies that, for node
$x$, if $y$ is the largest integer such that $2^y \le x+1$ and $z = x
- 2^y +1$, then:
\begin{itemize}
\item if $x\geq 2^b-1$, $x$ is a leaf;
\item $leftChild(x) = 2^{y+1}-1 + 2*z$ if $x$ is not a leaf;
\item $rightChild(x) = 2^{y+1}-1 + 2*z+1$ if $x$ is not a leaf;
\item $parent(x) = -1$ if $x=0$ ($x$ is the root);
\item $parent(x) = 2^{y-1}-1+|\frac{z}{2}|$ if $x\neq 0$;
\item $rightmostChild(x) = 2^b-1 +  (z+1)2^{b-z}  -1$.
\end{itemize}

\paragraph{Implementing the DP and LS algorithms with the hybrid data structure.}
It is then easy to implement the DP algorithm using the data structure
$B$, by using $\mathcal{B}$ as the current column of the DP table
(\emph{i.e.} if the currently processed position of $t'$ is $i$,
$\mathcal{B}[j]$ is the score of the best partial chain included in
the rectangle defined by $(0,0)$ and $(i,j)$), without updating the
internal nodes of the binary search tree $\mathcal{C}$. 

To implement the LS algorithm, the key points are 
\begin{itemize}
\item to be able to update efficiently the data structure $B$, when a
  fragment $s$ has been processed;
\item to be able to find  the best score of a partial chain
ending up at a position in $u'$ strictly below $p$.
\end{itemize}

Updating $B$ can be done through the function $setScore$ below, with
parameters $p=s.t$ and $score=S[s]$, while the second task can be
achieved by the function $getBestScore$ described below, which is a
simple binary tree search.

\begin{algorithm}\small
\caption{Set a chaining score for a position $p$.}
\label{algo:bt_insert}
\begin{lstlisting}[]
$setScore(B,p,score):$
   $index=2^b-1+p$ // start from leaf corresponding to $p$
   while $index!=-1$ && $B[index]<score$
      $B[index]=score$
      $index=parent(index)$
\end{lstlisting}
\end{algorithm}

\begin{algorithm}\small
\caption{Retrieve the best chaining score for partial chains ending strictly below position $p$.}
\label{algo:bt_find}
\begin{lstlisting}[]
$getBestScore(B,p):$
   let $b$ be the smallest integer s.t. $m'\leq 2^b$ 
   $maxScore=0$
   $currentNode=0$ // the root node
   $indexOfP=2^b-1+p$ 
   while $rightmostChild(currentNode)>indexOfP$
      $left = leftChild(currentNode)$
      if $rightmostChild(left)>=indexOfP$ // move left
         $ncurrentNode=left$
      else // move right
         $maxScore = max(maxScore,B[left])$
         $currentNode=rightChild(currentNode)$
   return $\max(maxScore,B[currentNode])$
\end{lstlisting}
\end{algorithm}

It is straightforward to see that if all updates of $B$ are done using
the function $setScore$, then the two required invariants on $B$ are
satisfied. The time complexity of both $setScore$ and $getBestScore$
is in $O(\log(m'))$, due to the fact that the binary tree is balanced.
So now, we can implement the LS algorithm on compact instances using
the data structure $B$ by replacing the instruction in line
\ref{algo:ls:use_A} of the LS algorithm by a call to
$getBestScore(B,s.b)$, the block of instructions in lines
\ref{algo:ls:A_1}-\ref{algo:ls:A_2} by $setScore(B,S[s])$ and reading
the optimal chain score in the root of the binary tree. The complexity
of operations over $B$ are logarithmic in $m'$ that is less or equal
to $k$. Thus the overall time complexity is in $O(k\log m')$.

\paragraph{LS/DP update with the hybrid data structure.}
So, in an hybrid algorithm that relies on the data structure $B$, when
the algorithm switches approaches (from DP to LS, or LS to DP), the
data structure $B$ is assumed to be consistent for the current
approach, and needs to be updated to become consistent for the next
approach.

So when switching from DP (say position $i-1$, $i=1,\dots n'$) to LS
(position $i$), we assume that $\mathcal{B}[j]$ ($j=0,\dots,m'-1$) is
the optimal score of a partial chain in the rectangle defined by
$(0,0)$ and $(i-1,j)$, and we want to update $B$ in such a way that
the label of any internal node $x$ of the binary tree is the maximum
of both its children. As $\mathcal{B}$ are the leaves of the binary
tree, this update can be done during a post-order traversal of the
binary tree, so in time $O(m')$.

When switching from LS to DP (say to use the DP approach on position
$i$ while the LS approach was used on position $i-1$), we assume that
for every leaf $\mathcal{B}[j]$ of the binary tree corresponding to a
position at most $i-1$, the value in $\mathcal{B}[j]$ is the optimal
score of a partial chain in the whose last fragment ends in position
$i-1$; this follows immediately from the way labels of the leaves of
the binary tree are inserted by the $setScore$ function. To update
$B$, we want that in fact $\mathcal{B}[j]$ is the optimal score of a
partial chain in the whose last fragment ends in position at most
$i-1$. So the update function needs only to give to $\mathcal{B}[j]$
the value $\max_{0\leq j'\leq j}\mathcal{B}[j']$, which can again be
done in time $O(m')$.

So updating the data structure $B$ from DP to LS or LS to DP
can be done in time $O(m')$. We denote by $update$ the function
performing this update.

\paragraph{Deciding between LS and DP using the fragment density.}
Before we can finally introduce our algorithm, we need to address the
key point of how to decide which paradigm (DP or LS) to use when
processing the fragments having a border in the current position of $t$,
say $c$.  Let $\mathcal{K}_c$ be the number of fragments $s$ such that
$s.\ell=c$ or $s.r=c$.  Using the DP approach, the cost of updating
$\mathcal{B}$ (\emph{i.e.} to compute the column $c$ of the DP table) is
$O(m'+\mathcal{K}_c)$. With the LS approach, the cost of updating $B$ is
in $O(\mathcal{K}_c\log {m'})$.

So, if $\mathcal{K}_c > \frac{m'}{\log m' -1}$, the asymptotic cost of
the DP approach is better than the asymptotic cost of the LS approach,
while it is the converse if $\mathcal{K}_c \leq \frac{m'}{\log m'
  -1}$. So, prior to processing fragments, for each position $i$ in $t$
($i=0,\dots,m'-1$), we record in an array $C$ is fragments borders in
position $i$ are processed using the DP approach ($C[i]$ contains DP)
or the LS approach ($C[i]$ contains LS). This last observation leads
to our main result, Algorithm~\ref{algo:hybrid} below.

\begin{algorithm}\small
\caption{A hybrid algorithm for the fragment chaining problem.}
\label{algo:hybrid}
\begin{lstlisting}[]
compute the compact instance $(t',u',{\cal S}')$
$L1$: an array of $n'\times 2$ linked lists
$C$: an binary array of size $n'$
foreach $s$ in ${\cal S}'$ do (*@\label{algo:hybrid:border1}@*)
   if $C[s.r]$ is DP then front insert $(s,end,s.t)$ into  $L1[s.r][1]$   
   else  front insert $(s,end)$ into  $L1[s.r][0]$
   if $C[s.{\ell}]$ is DP then front insert $(s,begin,s.b)$ into $L1[s.{\ell}][1]$
   else front insert $(s,begin)$ into $L1[s.{\ell}][0]$ (*@\label{algo:hybrid:border2}@*)
$B$: a binary tree for $m'$ leafs (all nodes are set to zero)
$\mathcal{B}$: refers to the $m'$ leaves of $B$
$S$: an array of integer of size $k$
for $i$ from $0$ to $n'$ do (*@\label{algo:hybrid:column_loop}@*)
   if $C[i] \neq C[i-1]$ then $update(B)$ (*@\label{algo:hybrid:update}@*)
   if $C[i]$ is DP  (*@\label{algo:hybrid:direct_loop1}@*)
      $L2$: an array of $m'$ linked lists
      for each $(s,t,j)$ in $L1[i][1]$ do front insert $(s,t)$ into  $L2[j]$
      $left = 0$, $leftDown = 0$
      for $j$ from $0$ to $m'$ do
         $maxC = 0$
         foreach $(s,type)$ in $L2[j]$ do
            if $type$ is $begin$ then $S[s] = s.s + leftDown$
            if $type$ is $end$ and $S[s]>maxC$ then $maxC = S[s]$
         $leftDown = left$, $left = \mathcal{B}[j]$
         $\mathcal{B}[j] = max(\mathcal{B}[j],\mathcal{B}[j-1],maxC)$ (*@\label{algo:hybrid:direct_loop2}@*)
   else // $C[i]$ is LS (*@\label{algo:hybrid:sweep_loop1}@*)
      foreach $(s,type)$ in $L1[i][0]$ do
          if $type$ is $begin$ then $S[s] = s.s + getBestScore(B,s.b)$
          if $type$ is $end$ then $setScore(B,s.t,S[s])$ (*@\label{algo:hybrid:sweep_loop2}@*)
if $C[n'-1]$ is $direct$ then return $\mathcal{B}[m'-1]$
else return value of the root of $B$
\end{lstlisting}
\end{algorithm}

\paragraph{Time and space complexity.}
In terms of space complexity, the algorithm, we avoid to use $O(k+
n'\times m')$ space for storing the fragments borders in $n'\times m'$
lists (structure $L$ of the DP algorithm) by using two lists: $L1[i]$
stores all fragments borders in position $i$ of $t'$, while $L2[j]$ stores
all fragments borders in position $i$ of $t'$ and $j$ of $u'$, and is
computed from $L[1]$. So the total space requirement is in
$O(k+m'+n')$.

We now establish the time complexity of this algorithm.  If the
current position $i$ of $t$ is tagged as DP, the cost for updating
the column is $O(m' + \mathcal{K}_i)$, including the cost of setting
up $L2$ from $L1$, that is proportional to the number of fragments borders
in the current position (line
\ref{algo:hybrid:direct_loop1}--\ref{algo:hybrid:direct_loop2}).  If
$C[i]$ is LS, the cost for computing chains scores on this position
is $O(\mathcal{K}_i\log m')$ (line \ref{algo:hybrid:sweep_loop1}--
\ref{algo:hybrid:sweep_loop2}). Thus, if we call $P^1$ the set of
positions on $t$ where we use the DP approach, $P^2$ the set of
positions on $t$ where we use the LS approach and $P=P^1 \cup P^2$,
the time for the whole loop at line \ref{algo:hybrid:column_loop}
is
$$
O\left( \sum_{p \in P^1} ( m' + \mathcal{K}_p ) + \sum_{p \in P^2} \mathcal{K}_p\log m' \right) 
$$ We have $|P^1| + |P^2| = n'$, $\forall p \in P^1: \mathcal{K}_p >
\frac{m'}{\log m' -1}$ and $\forall p \in P^2: \mathcal{K}_p \leq
\frac{m'}{\log m' -1}$. Moreover, updating the data structure $B$ from
LS to DP or DP to LS (line \ref{algo:hybrid:update}) is done at most
one more time then the size of $P^1$, so the total cost of this
operation is $O\left( \sum_{p \in P^1} m'\right)$, and can thus be
integrated, asymptotically, to the cost of processing the positions in
$P^1$.

\begin{theorem}\label{thm:main}
  The hybrid algorithm computes an optimal chain score in time 
  \begin{equation}O\left( k + \min(k\log k,m) + \min(k\log k,n) + \sum_{p \in P^1} (
  m' + \mathcal{K}_p ) + \log m' \sum_{p \in P^2}
  \mathcal{K}_p\right) \label{form:hybrid_complexity}
\end{equation} and space $O(k+n+m)$.
\end{theorem}

To conclude the complexity analysis, we show that the hybrid algorithm
performs at least as well, asymptotically, than both the DP and the LS
algorithms.  From (\ref{form:hybrid_complexity}), we deduce that, if
$P^2=P$, the hybrid algorithm time complexity becomes $ O( k +
\min(k\log k,m) + \min(k\log k,n) + \log m' k) $, which is at worst
equal to the asymptotic worst-case time complexity of the LS algorithm
as $m'=\min(m,k)$.

Now, if $P^2\neq P$, for every position $c$ in $P^1$, we know that the
cost of updating $B$ and processing $c$ with the DP approach is not
worse than processing it with the LS approach, by the value chosen for
$\mathcal{K}_c$. This ensures that, asymptotically, the hybrid
algorithm does perform at least as well as the LS algorithm.



We consider now the DP algorithm. Again, from
(\ref{form:hybrid_complexity}), if $P^1=P$, the complexity becomes $
O( k + \min(k\log k,m) + \min(k\log k,n) + m'n') $, which is equal to
the original dynamic programming algorithm time complexity as
$n'=\min(n,k)$ and $m'=\min(m,k)$.

As above, if we assume now that $P^1\neq P$, then we know that the
cost of processing the positions of $P^2$ with the LS approach is
asymptotically not worse than processing them with the DP
algorithm. The cost of updating $B$ from switching from DP to LS can
be integrated into the asymptotic cost of the DP part. This shows that
the hybrid algorithm is, asymptotically, not worse than the pure DP
algorithm.

\section{Discussion}\label{sec:conc}


Our main result in the present paper is an hybrid algorithm that
combines the positive features of both the classical dynamic
programming and of the line sweep algorithm for the fragment chaining
problem. We did show that a simple data structure can be used to
alternate between both algorithmic principles, thus benefiting of the
positive behavior of both algorithms. Not surprisingly, the choice
between using the DP or the LS principle is based on fragments
density.

It is easy to define instances where the hybrid algorithm performs
better , asymptotically, than both the DP and LS algorithms. For
example, if $m=n^{\frac{4}{5}}$ and $k=2n^{\frac{3}{2}}$ and there are
$n^{\frac{3}{2}}$ seeds extremities on $t[0]$ and $n^{\frac{3}{2}}$
extremities on $t[n-1]$, all other extremities spread along $t$ and
$u$, we can show that the complexities are $O(n^{\frac{9}{5}})$ for
the DP algorithm, $O(n^{\frac{3}{2}}\log n)$ for the LS algorithm and
$O(n^{\frac{3}{2}})$ for the hybrid algorithm.  However, so far our
result is mostly theoretical. The threshold of $m'/(\log(m')-1)$
considered on real genome data is high, as it assumes a very high
vfragment density that is unlikely to be observed often, at least on
applications such the alignment of whole bacterial genomes for
example. Preliminary experiments on such dfatya following t5he
approach developped in~\cite{UricaruMR11} show that the LS algorithm
is slightly more efficient than the hybrid one.  So it remains to be
seen if it could result in an effective speed-up when chaining
fragments in actual biological applications, especially involving
high-throughput sequencing data or overlapping
fragments~\cite{UricaruMR11}. From a practical point of view, it is
also of interest to consider algorithm engineering apsects, especially
related to the hybrid data structure, to see if this could alleviate
the issue of the high density threshold required to switch between the
LS and DP approaches, and assess the practical interest of the novel
theoretical framework we introduced in the present paper.

\end{document}